\documentstyle[11pt,aaspp4]{article}

\newcommand{\ubf}{{\bf u}}
\newcommand{\vbf}{{\bf v}}
\newcommand{\wbf}{{\bf w}}
\newcommand{\rbf}{{\bf r}}
\newcommand{\Omegabf}{{\bf \Omega}}

\begin{document}

\title{Irrotational Binary Neutron Stars\\
 in Quasiequilibrium in General Relativity}

\author{Saul A. Teukolsky\altaffilmark{1}}
\affil{Center for Radiophysics and Space Research, Cornell University,
Ithaca, NY 14853.}

\altaffiltext{1}{Also Departments of Physics and Astronomy, Cornell University}

\begin{abstract}
Neutron stars in binary orbit emit gravitational waves and spiral slowly
together. During this inspiral, they are expected
to have very little vorticity.
It is in fact a good approximation to treat
the system as having zero vorticity,
i.e., as irrotational.
Because the orbital period is much shorter than the radiation reaction
time scale, it is also an excellent approximation to treat the system
as evolving through a sequence of equilibrium states, in each of
which the gravitational radiation is neglected.
In Newtonian gravity, one can simplify the hydrodynamic equations
considerably for an equilibrium irrotational binary by introducing
a velocity potential. The equations reduce to a Poisson-like equation
for the potential, and a Bernoulli-type integral for the density. We show
that a similar simplification can be carried out in general relativity.
The resulting equations are much easier to solve than other
formulations of the problem.
\end{abstract}


\section{Introduction}

Ever since the discovery of the first neutron
star binary by Hulse \& Taylor (1975),
we have known that these systems exist
and will undergo orbital decay via gravitational wave emission. 
The inspiral and coalescence of binary neutron stars is one of the
primary targets for gravitational wave detectors
now under construction, like LIGO, VIRGO, and GEO (see, e.g., \cite{abr92},
or \cite{tho94}). And the coalescence
of binary neutron stars is the basis for several models for
gamma-ray bursters (\cite{pac86}; \cite{eich89}; \cite{nar92}).
Clearly, a theoretical understanding of coalescing binary neutron stars
is an important astrophysical problem. 

Analyzing coalescing neutron stars requires the full machinery
of general relativity. While some issues can be addressed with
Newtonian or post-Newtonian
gravity, such as the gravitational wave form at large separation,
others cannot even be posed in these weak-field limits. 
For example, consider the recent controversial claim by Wilson,
Mathews, and Marronetti (\cite{wil95}; \cite{wil96}; \cite{mat97}) that
massive neutron stars collapse to black holes before they merge.
Gravitational collapse of a star in equilibrium
is a consequence of the nonlinear nature of Einstein's equations;
it cannot be treated correctly with Newtonian gravity.
If neutron star binaries do in fact collapse before they merge,
there are enormous implications for gravitational wave detection:
The entire detection strategy is based on having accurate templates for
the wave forms, and current templates predict relatively high
sensitivity because without collapse
there are many orbital periods before coalescence.
And of course if the neutron stars collapse, then their
final coalescence cannot be a source of gamma rays.

There are other aspects of the inspiral problem that require a fully
relativistic treatment. For example, if the stars do not collapse before
merger, their combined mass is likely to be greater than the
maximum mass of a cold, rotating neutron star. Then the merged remnant
must ultimately collapse to a black hole. But it is not clear
whether this collapse occurs immediately on a dynamical time scale,
or whether thermal pressure from shock
heating will be sufficient to hold the
star up for a while. In this case, collapse will occur on a
neutrino-dissipation time scale. We also do not know how
much angular momentum the final remnant will have. What will the fate
of the system be if the angular momentum exceeds the maximum allowed
value for a Kerr black hole? Will the excess angular momentum be radiated
away gravitationally, or will it be ejected in a circumstellar ring of
matter, possibly leading to planet formation? All these questions
have observational implications that can be addressed only by fully
relativistic simulations.

A number of groups have undertaken the construction of general
relativistic codes that can treat the binary problem. One needs a code
that solves Einstein's equations in three spatial dimensions plus time,
and simultaneously solves the equations of relativistic hydrodynamics.
This is a formidable challenge, and not surprisingly various
approximations to the full problem have been proposed.
As we have argued above, it seems important not to give up the strong-field
aspects of the problem by using Newtonian or post-Newtonian gravity.
Instead, we will focus on two approximations that are likely to be
well-satisfied for real neutron star binaries, at least up until
the orbit becomes unstable and the stars finally plunge
together: {\em quasiequilibrium} and {\em irrotational flow}.

In Newtonian gravity, one can find exact equilibrium states for a
binary neutron star system. In general relativity, a binary loses
energy by gravitational wave emission and the stars spiral ever closer
together. However,
the time scale on which the orbit changes
is much longer than the orbital period, at least until the orbit becomes
unstable at the innermost stable circular orbit. We can therefore approximate
the evolution as proceeding through a succession of equilibrium states
of decreasing separation,
in each of which gravitational radiation can be neglected. The rate
of progression along the sequence is determined by the rate of gravitational
wave emission, but the structure at each point along the sequence can
be calculated to excellent approximation ignoring the radiation.

Relativistic quasiequilibrium neutron star binaries have
been constructed by
Baumgarte et al.\ (1997a,b,c). For Einstein's equations, they used
the approximation scheme of Wilson and Mathews (\cite{wil89}; \cite{wil90};
\cite{wil95}), which is a consistent way of neglecting small gravitational
radiation terms in Einstein's equations. This scheme has been calibrated
by Cook, Shapiro, \& Teukolsky (1996) on rapidly rotating single
neutron stars, where it gives
excellent results. Wilson and Mathews evolve
the binary system by solving the full dynamical equations for the
matter in the instantaneous background metric, and then updating the
metric quantities at each time step by re-solving the approximate
Einstein's equations. Baumgarte et al., on the other
hand, take advantage of the
quasiequilibrium approximation for the matter equations too: They
reduce the matter equations to a Bernoulli integral, which can be
solved much more accurately and conveniently than the original
equations of motion.

Using this approach Baumgarte et al.\ (1997d) studied the stability of
binary neutron stars and found no evidence for the instability
reported by Wilson and Mathews, even though they were using a very
similar approximation scheme. One possible reason for the difference
is that the models of Baumgarte et al. are corotating, that is,
are locked in
synchronous orbit. The models of Wilson and Mathews, by contrast,
have very little intrinsic spin. It is possible in
principle that the higher spin rate of the corotating models
suppresses the instability (\cite{mat97}). As we shall see, the
irrotational approximation allows the construction of models
almost as easily as the corotating assumption does. However, the
irrotational models are essentially nonspinning and should allow
one to settle the collapse question definitively.

Realistic neutron stars are in fact unlikely to be corotating---the
viscosity is probably too low to synchronize the spin and orbital
angular velocities (\cite{bil92}; \cite{koc92};
\cite{lai94}). Instead, the stars are likely to have very low
intrinsic spins compared with the orbital frequency,
and to good approximation we can assume they are
irrotational (zero vorticity as seen from the inertial
frame). Thus besides being of use in
determining whether the Wilson-Mathews instability is real or not,
the irrotational approximation will produce models that are close
to those expected to actually occur.

Ultimately, even if neutron star binaries do not collapse prematurely
to black holes, they will reach an instability corresponding to the
innermost stable circular orbit.
The subsequent plunge and merger cannot be treated by the
quasiequilibrium approximation. However,
models constructed with the approximations in this paper should provide
excellent initial data for the fully dynamical codes
required to treat the final plunge. In fact, because of the extreme
computational requirements of dynamical codes, it is unlikely that they
will be able to follow the evolution from large separation at all.
Realistic initial data such as provided by the methods here
will be crucial.

\section{Irrotational Binaries in Newtonian Gravity}

Exact irrotational binaries in Newtonian
gravity were first constructed by Bonazzola et al.\ (1992).
They used the following method: Start with the equation of motion
for a  fluid in a uniformly rotating frame of reference:
\begin{equation}\label{eqmot}
{\partial \vbf \over \partial t'}+(\vbf\cdot\nabla)\vbf+2\Omegabf\times\vbf
+\Omegabf\times(\Omegabf\times\rbf)=-{1\over\rho_0}\nabla P-\nabla\phi.
\end{equation}
Here $\Omegabf$ is the constant orbital
angular velocity, $P$ is the pressure,
$\rho_0$ the density, and $\phi$ the Newtonian potential. The
origin of the coordinates is at the center of mass of the binary
system. The velocity
$\ubf$ in the inertial frame is related to the velocity in the rotating
frame by
\begin{equation}
\ubf=\Omegabf\times\rbf+\vbf.
\end{equation}
For later comparison with the relativistic case, note that the time
derivatives in the rotating and inertial frames are related by
\begin{equation}\label{lie}
{\partial\over\partial t'}={\partial\over\partial t}+{\cal L}_{\bf w},
\end{equation}
where $\wbf=\Omegabf\times\rbf$ is the orbital velocity and
${\cal L}$ denotes the Lie derivative.

Now assume stationarity in the rotating frame: $\partial /\partial t'\to
0$. Then equation (\ref{eqmot}) can be rewritten as
\begin{equation}\label{roteq}
\nabla(\case{1}{2}v^2+h_N+\phi+\phi_c)+[(\nabla\times\vbf)+
2\Omegabf]\times\vbf=0.
\end{equation}
Here
\begin{equation}
h_N=\int{dP\over\rho_0}=u+P/\rho_0
\end{equation}
is the Newtonian enthalpy per unit mass, $u$ is the internal energy per
unit mass, and
\begin{equation}
\phi_c=-\frac{1}{2}(\Omegabf\times\rbf)^2
\end{equation}
is the centrifugal potential. Equation (\ref{roteq}) can also be written
in terms of the inertial velocity:
\begin{equation}\label{inerteq}
\nabla(\case{1}{2}u^2-(\Omegabf\times\rbf)\cdot\ubf+h_N+\phi)
+(\nabla\times\ubf)\times(\ubf-\Omegabf\times\rbf)=0.
\end{equation}
The continuity equation, with $\partial /\partial t'\to 0$, becomes
\begin{equation}\label{rotcont}
\nabla\cdot\vbf=-\vbf\cdot{\nabla\rho_0\over\rho_0},
\end{equation}
or
\begin{equation}\label{inertcont}
\nabla\cdot\ubf=-(\ubf-\Omegabf\times\rbf)\cdot{\nabla\rho_0\over\rho_0}.
\end{equation}

The corotating case corresponds to $\vbf=0$. Equation (\ref{rotcont})
is trivially satisfied, while equation (\ref{roteq}) is just the
Bernoulli equation:
\begin{equation}
h_N+\phi+\phi_c={\rm const}.
\end{equation}
The irrotational case corresponds to no vorticity in the inertial
frame, $\nabla\times\ubf=0$. This implies that $\ubf$ is given by
a velocity potential,
\begin{equation}
\ubf=\nabla\psi.
\end{equation}
Equations (\ref{inerteq}) and (\ref{inertcont}) become
\begin{equation}\label{erig1}
\case{1}{2}(\nabla\psi)^2-(\Omegabf\times\rbf)\cdot\nabla\psi
+h_N+\phi={\rm const},
\end{equation}
\begin{equation}\label{erig2}
\nabla^2\psi=-(\nabla\psi-\Omegabf\times\rbf)\cdot{\nabla\rho_0\over
\rho_0}.
\end{equation}
Equation (\ref{erig2}) must be solved subject to the boundary condition
\begin{equation}
(\nabla\psi-\Omegabf\times\rbf)\cdot\nabla\rho_0|_{\rm surf}=0,
\end{equation}
where the surface is defined by $\rho_0=0$.

Note that in terms of the velocity in the rotating frame, $\vbf$,
its divergence is given by equation
(\ref{rotcont}) while its curl is
\begin{equation}\label{curlv}
\nabla\times\vbf=-2\Omegabf.
\end{equation}
The Bernoulli integral is
\begin{equation}\label{rotbern}
\case{1}{2}v^2+h_N+\phi+\phi_c=\rm const.
\end{equation}
The curl equation (\ref{curlv}) is solved by
\begin{equation}\label{decompv}
\vbf=-\Omegabf\times\rbf+\nabla\psi,
\end{equation}
and we then recover equations (\ref{erig1}) and (\ref{erig2}) from
equations (\ref{rotbern}) and (\ref{rotcont}).

Equations (\ref{erig1}) and (\ref{erig2}) were solved by
Bonazzola et al.\ (1992) for an illustrative case, but they did
not give any sequences of models. The first sequences
were constructed by Ury\=u \& Eriguchi (1997), who
solved the same equations for equal mass
binaries with polytropic equations of state.

\section{Irrotational Flow in General Relativity}

Just as in Newtonian fluid mechanics, the velocity of a relativistic
perfect fluid can be expressed as the gradient of a potential if the
vorticity is zero (see, e.g., \cite{lan59}, or \cite{mon80}).
Define the relativistic enthalpy by
\begin{equation}\label{h}
h={\rho+P\over \rho_0},
\end{equation}
where $\rho$ is the total energy density
and $\rho_0=m_Bn$ is the rest-mass density. (We use units with
$c=G=1$.) Here $n$ is the baryon number
density and $m_B$ the mean baryon rest mass. Then
the relativistic vorticity tensor is defined as
\begin{equation}\label{vort}
\omega_{\mu\nu}=P^\alpha{}_\mu P^\beta{}_\nu[\nabla_\beta(hu_\alpha)
-\nabla_\alpha(hu_\beta)].
\end{equation}
Here $u^\mu$ is the fluid 4-velocity and $P^\mu{}_\nu=\delta^\mu{}_\nu
+u^\mu u_\nu$ is the projection tensor.
There is another definition of relativistic vorticity, without the
$h$ in equation (\ref{vort}). Both definitions reduce to the correct
Newtonian limit, where $h\to 1$, but the definition (\ref{vort}) is the
one that leads to a natural definition of potential flow.

For a perfect fluid, Euler's equation can be written
\begin{equation}\label{euler}
u^\alpha\nabla_\alpha(hu_\mu)+\nabla_\mu h=0.
\end{equation}
Equations (\ref{vort}) and (\ref{euler}) yield a simple expression for
the vorticity,
\begin{equation}
\omega_{\mu\nu}=\nabla_\nu(hu_\mu)-\nabla_\mu(hu_\nu).
\end{equation}
Thus if the vorticity is zero, then the quantity $hu_u$ can be expressed
as the gradient of a potential:
\begin{equation}\label{psidef}
hu_\mu=\nabla_\mu\psi.
\end{equation}

The equation of continuity for the baryon density $n$ is
\begin{equation}
\nabla_\alpha(nu^\alpha)=0,
\end{equation}
or,
\begin{equation}\label{nonlin}
\nabla^\alpha[(n/h)\nabla_\alpha\psi]=0.
\end{equation}
The equation of state relates $n$ to $h$, and $h$ is found from the
normalization condition $u_\alpha u^\alpha=-1$, which by equation
(\ref{psidef}) gives
\begin{equation}\label{hfrompsi}
h=[-(\nabla_\alpha\psi)(\nabla^\alpha\psi)]^{1/2}.
\end{equation}

Even for flow in a fixed background geometry, equation (\ref{nonlin})
with equation (\ref{hfrompsi}) is in general a nonlinear equation in
$\psi$ and its derivatives. We rewrite it as
\begin{equation}\label{psieqn}
\nabla^\alpha\nabla_\alpha\psi=-(\nabla^\alpha\psi)\nabla_\alpha\ln(n/h).
\end{equation}
Then one way of solving equation (\ref{psieqn}) is by iteration,
with either $n/h$ or the whole right-hand side determined from the
previous iteration. In general, this procedure will have to be iterated
with the solution of the Einstein field equations for the metric.

\section{3+1 Decomposition}

We make the usual ADM 3+1 decomposition of the spacetime metric:
\begin{equation}
ds^2=-\alpha^2 dt^2+\gamma_{ij}(dx^i+\beta^i dt)(dx^j+\beta^j dt),
\end{equation}
where $\alpha$ is the lapse, $\beta^i$ is the shift, and $\gamma_{ij}$
is the 3-metric on the $t={}$constant time slices. If we let $\vec n$
denote the unit normal to these time slices, and $\vec k$ the tangent
to the $t$-coordinate lines (i.e., $\vec k=\partial/\partial t$ in the
ADM coordinate system), the the decomposition is equivalent to writing
\begin{equation}\label{kdef}
\vec k = \alpha \vec n + \vec \beta, \qquad \vec n\cdot\vec\beta=0,
\end{equation}
and expressing the 3-metric in a general coordinate system as
\begin{equation}
\gamma_{\mu\nu}=g_{\mu\nu}+n_\mu n_\nu.
\end{equation}
We will call any object that is orthogonal to $\vec n$ {\em spatial}.

The spatial (or 3-dimensional) covariant derivative $D_\mu$ of any tensor is
defined by taking the 4-dimensional covariant derivative $\nabla_\mu$
and then projecting each free index with a $\gamma_{\mu\nu}$ so that the
result is completely spatial. Since $D_\mu\gamma_{\alpha\beta}=0$, this
covariant derivative is compatible with $\gamma_{\alpha\beta}$, and
since a compatible covariant derivative is unique, this definition is
equivalent to defining $D_\mu$ as the covariant derivative with respect
to the 3-metric $\gamma_{\alpha\beta}$.

For later reference, we list some standard relations that hold in the
3+1 decomposition. The derivative of the normal vector is
\begin{equation}\label{gradn}
\nabla_\mu n_\nu=-K_{\mu\nu}-n_\mu D_\nu\ln\alpha.
\end{equation}
Here $K_{\mu\nu}$ is the extrinsic curvature tensor, which is spatial.
From equation (\ref{gradn}) we see that
\begin{equation}\label{ngradn}
n^\mu\nabla_\mu n_\nu=D_\nu\ln\alpha.
\end{equation}
The definition of $D_\mu$ gives
\begin{equation}\label{gradf}
\nabla_\mu f=D_\mu f - n_\mu n^\nu\nabla_\nu f,
\end{equation}
where $f$ is any scalar. Similarly, for any spatial vector $\vec X$, we have
\begin{equation}\label{gradx}
\nabla_\mu X_\nu = D_\mu X_\nu - n_\mu n^\lambda\nabla_\lambda X_\nu
-K_{\mu\lambda}X^\lambda n_\nu.
\end{equation}

\section{Quasiequilibrium}

As discussed in the Introduction, it is a good approximation to treat
neutron star binaries as essentially equilibrium states, and to
neglect the gravitational radiation on orbital time scales.
The quasiequilibrium assumption is implemented mathematically by requiring
that the spacetime admit a Killing vector to express the symmetry: a
rotation $\Delta\phi$ about the rotation axis is equivalent to
a displacement $\Omega\Delta t$, where $\Omega$ is the uniform orbital
angular velocity. We write the Killing vector as
\begin{equation}
\vec l={\partial\over\partial t}+\Omega{\partial\over\partial\phi}\equiv
\vec k + \Omega\vec m.
\end{equation}
Here the generator of rotations about the rotation axis of the binary
is denoted by $\partial/\partial\phi$, or $\vec m$ in an arbitrary coordinate
system. By equation (\ref{kdef}), we can also write
\begin{equation}\label{ldef}
\vec l = \alpha \vec n +\vec B,
\end{equation}
where
\begin{equation}
\vec B=\vec \beta+\Omega\vec m
\end{equation}
is the rotating shift vector. It is equivalent to adding
$\bf \Omega\times r$ to the original shift. (Note that here $\vec B$ is
the negative of $\vec B$ defined by \cite{bona97}.) While $\vec l$ is not
a Killing vector in the exact metric, in the
quasiequilibrium approximation Einstein's equations can be
consistently approximated to neglect gravitational radiation and
admit $\vec l$ as a Killing vector, e.g., by the Wilson-Mathews scheme
(\cite{wil89}; \cite{wil90}; \cite{wil95}).

We require that $\vec l$ be a symmetry generator for the matter fields
as well as for the metric. Mathematically, this means that the Lie
derivative along $\vec l$ of any matter field must vanish (cf.\
eq.\ \ref{lie}). Note that
since $\psi$ is a potential, we must be careful that the symmetry
condition is expressed correctly for it. It is not true that
\begin{equation}
{\partial \psi \over \partial t}+\Omega{\partial\psi\over\partial\phi}
=0\qquad\hbox{(wrong!)}
\end{equation}
Rather,
\begin{equation}
0={\cal L}_{\vec l}(h\tilde u)={\cal L}_{\vec l}(d\psi)=d({\cal L}
_{\vec l}\psi).
\end{equation}
Here we have used equation (\ref{psidef}) in the notation of differential
forms, and used the fact that the Lie derivative of a form commutes with
the exterior derivative. Thus as long as ${\cal L}_{\vec l}\psi$ is
a constant the symmetry will be respected.
We write
\begin{equation}\label{cdef}
l^\mu\nabla_\mu\psi={\cal L}_{\vec l}\psi=
{\partial \psi \over \partial t}+\Omega{\partial\psi\over\partial\phi}
=-C,
\end{equation}
where the minus sign is chosen for later convenience:
$C$ is a positive constant.

Another way of seeing why equation (\ref{cdef}) holds
is to project the Euler equation (\ref{euler})
along $\vec l$. Using Killing's equation
\begin{equation}\label{killing}
\nabla_\mu l_\nu+\nabla_\nu l_\mu=0,
\end{equation}
and the fact that $h$ satisfies the
symmetry, we can derive the Bernoulli integral
\begin{equation}\label{bern}
hu_\mu l^\mu=\rm constant.
\end{equation}
By equation (\ref{psidef}), this is equivalent to equation (\ref{cdef}).

One way of implementing equation (\ref{cdef}) is to solve equation
(\ref{psieqn}) by separating out the $t$-dependence with a solution of
the form
\begin{equation}
\psi=-Ct+f(r,\theta,\phi-\Omega t).
\end{equation}
We will proceed instead to derive a coordinate-independent equation
that embodies the symmetry.

We will need expressions for the normal derivatives of scalar
quantities
that satisfy the symmetry. For $\psi$, equations (\ref{ldef}) and
(\ref{cdef}) give
\begin{eqnarray}\label{gradnpsi}
n^\mu\nabla_\mu\psi & = &{1\over \alpha}(l^\mu-B^\mu)\nabla_\mu\psi
\nonumber \\
& = &-{1\over\alpha}(C+B^\mu D_\mu \psi).
\end{eqnarray}
For a scalar $f$ that satisfies ${\cal L}_{\vec l}f=0$ we have a
similar equation with $C=0$:
\begin{equation}\label{gradnf}
n^\mu\nabla_\mu f=-{1\over\alpha}B^\mu D_\mu f.
\end{equation}

\section{3+1 Decomposition of the Potential Equation}

We can now derive the 3+1 decomposition of the potential equation
(\ref{psieqn}). First, equations (\ref{gradf}) and (\ref{gradnpsi})
give
\begin{equation}\label{gradpsi}
\nabla_\mu\psi=D_\mu\psi + n_\mu(C+B^\nu D_\nu\psi)/\alpha.
\end{equation}
Therefore
\begin{equation}\label{gradgradpsi}
\nabla^\mu\nabla_\mu\psi=\nabla^\mu D_\mu\psi+(\nabla^\mu n_\mu)
(C+B^\nu D_\nu\psi)/\alpha +n^\mu\nabla_\mu(C+B^\nu D_\nu\psi)/\alpha.
\end{equation}
Using equation (\ref{gradx}) with $X_\nu\to D_\nu\psi$, we can write
the first term on the right-hand side of (\ref{gradgradpsi}) as
\begin{eqnarray}\label{one}
\nabla^\mu D_\mu\psi & = & D^\mu D_\mu\psi -n^\mu n^\lambda \nabla_\lambda
D_\mu\psi \nonumber \\
& = & D^\mu D_\mu\psi -n^\lambda \nabla_\lambda (n^\mu D_\mu\psi)
+(n^\lambda \nabla_\lambda n^\mu)D_\mu\psi \nonumber \\
& = & D^\mu D_\mu\psi -0+ (D_\mu\psi) D^\mu\ln\alpha.
\end{eqnarray}
In the second term on the right-hand side of (\ref{gradgradpsi}) we
replace $\nabla^\mu n_\mu$ by $-K$ (${}=-K^\mu{}_\mu$). For the
third term, we use equation (\ref{gradnf}):
\begin{equation}\label{two}
n^\mu\nabla_\mu{1\over\alpha}(C+B^\nu D_\nu\psi) = -{1\over\alpha}
B^\mu D_\mu{1\over\alpha}(C+B^\nu D_\nu\psi).
\end{equation}
Substituting equations (\ref{one}) and (\ref{two}) in equation
(\ref{gradgradpsi}), we find
\begin{eqnarray}
\nabla^\mu\nabla_\mu\psi & = & D^\mu D_\mu\psi +(D_\mu\psi) D^\mu\ln\alpha
-{K\over\alpha}(C+B^\nu D_\nu\psi) \nonumber \\
& & {}+{1\over\alpha^3}(B^\mu D_\mu\alpha)
(C+B^\nu D_\nu\psi)-{B^\mu\over\alpha^2}D_\mu(B^\nu D_\nu\psi).
\end{eqnarray}

For the right-hand side of equation (\ref{psieqn}) we use equation
(\ref{gradpsi}) and a similar equation with $\psi$ replaced by $\ln(n/h)$
and $C=0$. So the final form of equation (\ref{psieqn}) is
\begin{eqnarray}\label{final}
D^i D_i\psi &+& (D_i\psi) D^i\ln\alpha
+(C+B^j D_j\psi)\left( {B^i\over\alpha^2}D_i\ln\alpha-{K\over\alpha}\right)
\nonumber \\
&-& {B^i\over\alpha^2}D_i(B^j D_j\psi)=-\left[D^i\psi
-(C+B^j D_j\psi){B^i\over\alpha^2}\right]D_i\ln\left({n\over h}
\right).
\end{eqnarray}
Here we have used latin indices since the equation is completely spatial,
with the metric being the 3-metric $\gamma_{ij}$. For numerical work,
it may be advantageous to rewrite equation (\ref{final}) in an equivalent
form. For example, define
\begin{equation}
\lambda=C+B^j D_j\psi=\alpha[h^2+(D_i\psi)^2]^{1/2},
\end{equation}
where the second equality follows from equation (\ref{cformula}) below.
Then equation (\ref{final}) becomes
\begin{equation}\label{final2}
D^i D_i\psi-B^iD_i{\lambda\over\alpha^2}-{\lambda K\over\alpha}=
- \left(D^i\psi
-{\lambda\over\alpha^2}B_i\right)D_i\ln\left({\alpha n\over h}
\right).
\end{equation}
An equivalent form is obtained by the substitution
\begin{equation}
\alpha K = D_i B^i,
\end{equation}
which follows from Killing's equation (\ref{killing}). Then
\begin{equation}\label{final3}
D^i D_i\psi-D_i{\lambda B^i\over\alpha^2}=
- \left(D^i\psi
-{\lambda\over\alpha^2}B_i\right)D_i\ln\left({\alpha n\over h}
\right).
\end{equation}
Since a consistent approximation scheme for Einstein's equations
with the assumptions made here, such as the
Wilson-Mathews scheme, typically has $K=0$, the form (\ref{final2})
may be better.

At the surface of either star, $n\to 0$, $h\to 1$.
Thus equation (\ref{final2}) or (\ref{final3})
must be solved together with the boundary condition
\begin{equation}
\left.\left(D^i\psi
-{\lambda\over\alpha^2}B_i\right)D_i n\right|_{\rm surf}=0.
\end{equation}

The Bernoulli integral for the matter distribution follows from
substituting equation (\ref{gradpsi}) in equation (\ref{hfrompsi}):
\begin{equation}\label{hformula}
h^2=-(D^i\psi) D_i\psi +{1\over\alpha^2}(C+B^j D_j\psi)^2,
\end{equation}
or equivalently
\begin{equation}\label{cformula}
C=-B^j D_j\psi+\alpha[h^2+(D_i\psi)^2]^{1/2}.
\end{equation}
The positive sign for the square root in equation (\ref{cformula}) is
determined by the Newtonian limit (cf.\ \S\ref{newtlim}).

\section{Newtonian Limit}\label{newtlim}

It is easy to see that the above equations reduce to the expected
form in the Newtonian limit. The Newtonian limit of the various
quantities is
\begin{eqnarray}
D_i\psi & \to & \partial_i\psi \\
B_i & \to & ({\bf \Omega\times r})_i \\
\alpha & \to & 1+\phi \\
h & \to & 1+h_N \\
K & \to & 0.
\end{eqnarray}
Thus equation (\ref{cformula}) becomes
\begin{equation}
C\to -({\bf \Omega\times r})_i\partial_i\psi+[1+\phi+h_N+\frac{1}{2}
(\partial_i\psi)^2].
\end{equation}
In leading order, $C\to 1$. If we define $C=1+C'$, then in next order
\begin{equation}
C'=\phi+h_N+\frac{1}{2}(\partial_i\psi)^2-
({\bf\Omega\times r})_i\partial_i\psi.
\end{equation}
This is equation (\ref{erig1}).

Similarly, neglecting $O(v^2)$ corrections, we see that equation
(\ref{final}) reduces to
\begin{equation}
\nabla^2\psi = -[\partial_i\psi -({\bf\Omega\times r})_i]
{\partial_i n \over n},
\end{equation}
which is equation (\ref{erig2}).

\section{Conclusions}

Equation (\ref{final2}), together with equation (\ref{hformula}) or
(\ref{cformula}), is the principal result of this paper. It is a
relatively simple Poisson-like equation for a scalar field that
determines the fluid velocity for irrotational flow
in the quasiequilibrium approximation. Bonazzola, Gourgoulhon, \&
Marck (1997) have considered the same problem as treated here,
but their formalism is considerably more complicated. Their results
are compared with ours in Appendix A. We do note, however, that this
paper was inspired by reading theirs.

The equations derived here, together with a compatible approximation
to Einstein's equations such as the Wilson-Mathews scheme, allow
one to construct sequences of neutron star binaries that should
be very good approximations to actual binaries during the inspiral
phase. The formalism is not much more complicated than that used
by Baumgarte et al.\ (1997a,b,c,d) to construct corotating sequences,
since equations very similar to equation (\ref{final}) are already
being solved in that case. Work is underway to construct such
more realistic sequences.

After this paper was submitted for publication, I became aware of
an independent paper by Shibata (1998) treating the same problem.
His equations (2.18) and (2.22) are completely equivalent to our
equations (\ref{cformula}) and (\ref{final3}) when the appropriate
substitutions are made.
\acknowledgments
I thank Larry Kidder for helpful discussions.
This work was supported in part by NSF Grant No.\ PHY 94-08378 to Cornell
University, and by the Grand Challenge Grant No.\ NSF PHY 93-18152/ASC
93-18152.

\appendix

\section{Comparison with the Formalism of Bonazzola, Gourgoulhon,
\& Marck}

Bonazzola, Gourgoulhon, \& Marck (1997) have developed a different
way of treating irrotational quasiequilibrium binaries.
Instead of working directly with the 4-velocity of the fluid
as we have, they consider the 3-velocity in the rotating frame.
They decompose the 4-velocity as
\begin{equation}
\vec u=\Gamma(\vec v + \vec V),
\end{equation}
where $\vec v \propto \vec l$
is the 4-velocity of the rotating frame, $\vec V$ is the
3-velocity in the rotating frame,
and $\Gamma=(1-\vec V\cdot\vec V)^{-1/2}$
is the Lorentz factor. Then they derive the Bernoulli integral
(\ref{bern}), which in their variables is
\begin{equation}
\ln h+\Phi+\ln\Gamma = \rm const,
\end{equation}
where $\Phi=\ln(\alpha^2-\vec B\cdot\vec B)^{1/2}$, and also equations
for the divergence and curl of the velocity field,
\begin{eqnarray}\label{divcurlv1}
\nabla\cdot\vec V & = & -\nabla_{\vec V}\ln(n\Gamma), \\ \label{divcurlv2}
\nabla\wedge\vec V & = & -2\vec \omega+(\nabla\wedge \vec V+2\vec \omega)\cdot\vec V
{\vec V\over \vec V\cdot\vec V}-\vec V \wedge\nabla\Phi.
\end{eqnarray}
The ``cross product'' operation $\wedge$ used
for the curl is defined in the rest frame
of the rotating observer:
\begin{equation}
(\vec a \wedge \vec b)_\mu=
v^\alpha\epsilon_{\alpha\mu\beta\gamma}a^\beta b^\gamma.
\end{equation}
The ``rotation vector'' $\vec \omega$ is defined by
\begin{equation}
\nabla\wedge\vec v = 2\vec \omega.
\end{equation}
In the Newtonian limit, where $\vec \omega \to \Omegabf$,
these equations reduce to equations (\ref{rotcont}),
(\ref{curlv}), and (\ref{rotbern}).

Bonazzola, Gourgoulhon, and Marck carry out the 3+1 decomposition of
the above equations 
to get 3-dimensional equations for the velocity field.
They then decompose the velocity using both scalar and vector
potentials, and obtain a scalar and a vector Poisson-like equation.
In the Newtonian limit, the vector potential has an analytic solution
corresponding to the term $-\Omegabf\times\rbf$ in equation (\ref{decompv}),
leaving just the scalar potential to be solved for. 
However, there does not seem to be an  analytic solution for
the vector potential in the relativistic case, and so one has
to solve for it numerically as well.

The procedure of Bonazzola, Gourgoulhon, \& Marck (1997) can actually
be simplified somewhat by working with a slightly different 3-dimensional
velocity $\vec p$, defined by
\begin{equation}
\vec p = e^{-\Phi}\vec V.
\end{equation}
Then equations (\ref{divcurlv1}) and (\ref{divcurlv2}) become
\begin{eqnarray}
\nabla\cdot\vec p & = & -\nabla_{\vec p}\ln(n\Gamma e^{\Phi}), \\
\nabla\wedge\vec p & = & -2e^{-\Phi}\vec \omega.
\end{eqnarray}
However, after the 3+1 decomposition and introduction of scalar and
vector potentials, one is still left with complicated equations.


\clearpage

\begin{thebibliography}{}
\bibitem[Abramovici et al.\ 1992]{abr92}
Abramovici, A., Althouse, W. E., Drever, R. W. P., Gursel, Y.,
Kawamura, S., Raab, F. J., Shoemaker, D., Sievers, L., Spero,
R. E., Thorne, K. S., Vogt, R. E., Weiss, R., Whitcomb, S. E., \&
Zucker, M. E. 1992,
Science, 256, 325
\bibitem[Baumgarte et al.\ 1997a]{baum97a} Baumgarte, T. W., Shapiro,
S. L., Cook, G. B., Scheel, M. A., \& Teukolsky, S. A. 1997a, in
Proceedings of the 18th Texas Symposium on Relativistic Astrophysics,
ed. A. Olinto, J. Friedman, \& D. Schramm (Singapore: World
Scientific)
\bibitem[Baumgarte et al.\ 1997b]{baum97b} Baumgarte, T. W.,
Cook, G. B.,  Scheel, M. A., Shapiro, S. L., \& Teukolsky, S. A. 1997b,
\prl, 79, 1182
\bibitem[Baumgarte et al.\ 1997c]{baum97c} Baumgarte, T. W.,
Cook, G. B.,  Scheel, M. A., Shapiro, S. L., \& Teukolsky, S. A. 1997c,
preprint gr-qc/9709026
\bibitem[Baumgarte et al.\ 1997d]{baum97d} Baumgarte, T. W.,
Cook, G. B.,  Scheel, M. A., Shapiro, S. L., \& Teukolsky, S. A. 1997d,
preprint gr-qc/9705023
\bibitem[Bildsten \& Cutler 1992]{bil92} Bildsten, L., \& Cutler,
C. 1992, \apj, 400, 175
\bibitem[Bonazzola et al.\ 1992]{bona92} Bonazzola,
S., Gourgoulhon, E., Haensel, P., \& Marck, J.-A. 1992, in Approaches
to Numerical Relativity, ed. R. d'Inverno (Cambridge: Cambridge
University Press), 230
\bibitem[Bonazzola, Gourgoulhon, \& Marck 1997]{bona97} Bonazzola,
S., Gourgoulhon, E., \& Marck, J.-A. 1997, \prd, 56, 7740
\bibitem[Cook, Shapiro, \& Teukolsky 1990]{cook90} Cook, G. B.,
Shapiro, S. L., \& Teukolsky, S. A. 1990, \prd, 53, 5533
\bibitem[Eichler et al.\ 1989]{eich89} Eichler, D., Livio, M., Piran, T.,
\& Schramm, D. N. 1989, Nature, 340, 126
\bibitem[Hulse \& Taylor 1975]{hulse} Hulse, R. A., \& Taylor,
J. H. 1975, \apjl, 195, L51
\bibitem[Kochanek 1992]{koc92} Kochanek, C. S. 1992, \apj, 398, 234
\bibitem[Lai 1994]{lai94} Lai, D. 1994, \mnras, 270, 611
\bibitem[Landau \& Lifshitz 1959]{lan59} Landau, L. D., \& Lifshitz, E. M.
1959, Fluid Mechanics (Reading: Addison-Wesley), 504
\bibitem[Mathews, Marronetti, \& Wilson 1997]{mat97} Mathews, G. J.,
Marronetti, P., \& Wilson, J. R. 1997, preprint gr-qc/9710140
\bibitem[Moncrief 1980]{mon80} Moncrief, V. 1980, \apj, 235, 1038
\bibitem[Narayan, Paczy\'nski, \& Piran 1992]{nar92} Narayan, R., Paczy\'nski,
B., \& Piran, T. 1992, \apjl, 395, L83
\bibitem[Paczy\'nski 1986]{pac86} Paczy\'nski, B. 1986, \apjl, 308, L51
\bibitem[Shibata 1998]{shib98} Shibata, M. 1998, preprint gr-qc/9803085.
\bibitem[Thorne 1994]{tho94} Thorne, K. S. 1994, in Relativistic
Cosmology, Proceedings of the 8th Nishinomiya-Yukawa Memorial Symposium,
ed. M. Sasaki (Tokyo: Universal Academy Press), 67
\bibitem[Ury\=u \& Eriguchi 1997]{uryu97} Ury\=u, K., \& Eriguchi,
Y. 1997, preprint astro-ph/9712203, submitted to \mnras
\bibitem[Wilson 1990]{wil90} Wilson, J. R. 1990, in Texas Symposium on
3-Dimensional Numerical Relativity, ed. R. A. Matzner (Austin: University
of Texas)
\bibitem[Wilson \& Mathews 1989]{wil89} Wilson, J. R., \& Mathews, G. J.
1989, in Frontiers in Numerical Relativity, ed. C. R. Evans, L. S. Finn,
\& D. W. Hobill (Cambridge: Cambridge University Press), 306
\bibitem[Wilson \& Mathews 1995]{wil95} Wilson, J. R., \& Mathews, G. J.
1995, \prl, 75, 4161
\bibitem[Wilson, Mathews, \& Marronetti 1996]{wil96} Wilson, J. R.,
Mathews, G. J., \& Marronetti, P. 1996, \prd, 54, 1317
\end{thebibliography}
\end{document}